\documentclass[prb,twocolumn,showpacs,nobalancelastpage,floatfix,preprintnumbers]{revtex4}

\preprint{\fbox{\texttt{\tiny\jobname.tex}}}

\usepackage{amsmath,amssymb,amsfonts}
\usepackage{bm,graphicx,color}

\newcommand{\om}{\omega}
\newcommand{\cd}{c^{\dagger}}
\newcommand{\cpd}{c^{\phantom\dagger}}
\newcommand{\la}{\langle}
\newcommand{\ra}{\rangle}
\newcommand{\gi}{\Gamma^{\infty}}
\newcommand{\tc}{\Tilde{\chi}}
\newcommand{\mbq}{\mathbf{q}}
\newcommand{\mbk}{\mathbf{k}}
\newcommand{\mbR}{\mathbf{R}}
\newcommand{\mbc}{\boldsymbol{\chi}}

\begin{document}

\title{Efficient treatment of two-particle vertices in dynamical mean-field theory}

\author{Jan Kune\v{s}}
\affiliation{Institute of Physics,
Academy of Sciences of the Czech Republic, Cukrovarnick\'a 10,
162 53 Praha 6, Czech Republic}

\date{\today}

\begin{abstract}
We present an efficient and numerically stable algorithm for calculation
of two-particle response functions within the dynamical mean-field theory.
The technique is based on inferring the high frequency asymptotic behavior of 
the irreducible vertex function from the local dynamical susceptibility.
The algorithm is tested on several examples. 
In all cases rapid convergence of the vertex function towards its
asymptotic form is observed.
\end{abstract}


\maketitle

\section{Introduction}
Electronic correlations in materials have been one of the central topics
of the condensed matter physics throughout its history encompassing
topics such as high-temperature superconductivity, colossal magneto-resistance
or heavy fermion physics. The introduction of dynamical mean-field theory (DMFT)
in early 1990's \cite{metzner89,georges92,rmp} marked a big step forward
in the theory of correlated electrons in providing an approximate, but
non-perturbative, computational method with several exact limits. Moreover, numerical DMFT calculations
are feasible also for multi-band Hamiltonians necessary for the description
of real materials. 

In the past twenty years DMFT was applied, at first, 
to models \cite{rmp}  and, later, to real materials \cite{rmp2,held07}. 
Naturally, most of the studies focused on
single-particle quantities, featured explicitly in the DMFT equations, which can 
be compared to photoemission spectra and which can provide information about the metal-insulator
transitions. Also local two-particle correlation functions, static 
as well as dynamical, can be computed with little additional effort
in most DMFT implementations, providing information about the local response
to applied fields. Computation of the non-local response is 
a more difficult venture, however, with great potential gains.
With the dynamical susceptibilities available it is possible to compare to the
experimental data from inelastic neutron, x-ray  
or electron loss spectroscopies. Perhaps even more interesting possibility
opens with the static susceptibilities. Monitoring
their divergencies as a function of temperature and the reciprocal lattice vector
allows investigation of the second order phase transitions and an unbiased determination
of the order parameters.
So far only calculations
for simple models, such as the single-band Hubbard
model \cite{jarrell92,ulmke95},the periodic Anderson model \cite{jarrell95,
fishman03,yu08}, or Holstein model \cite{freericks93}, have been reported. Similar
calculations were also performed with cluster \cite{maier05} 
or diagrammatic \cite{brener08} extentions of DMFT.

In this paper, we present a scheme for computation of the static two-particle
response functions in the multi-band Hubbard model within DMFT. The
key development is splitting the particle-hole irreducible vertex into
low frequency (LF) and high frequency (HF) parts, and 
expressing the HF part in terms of the local dynamical susceptibilities.
This allows reformulation of the Bethe-Salpeter equation in 
terms of the LF quantities plus corrections, thus reducing the numerical
cost of the calculations and improving their stability.
The procedure is applied to the single-band Hubbard model at and away from the half 
filling and a two-band bilayer model.
In all the studied cases we find a rapid convergence of 
the vertex function towards its asymptotic form, which leaves the size of the
LF problem rather small and manageable also for multi-band systems.

The paper is organized as follows. After a general introduction, the two-particle
formalism is reviewed in section II, followed by the discussion of the numerical 
implementation. In section III applications to simple model systems
are reported. Discussion of the asymptotic behavior
of the irreducible vertex and the blocked form
of the irreducible Bethe-Saltpeter equation is left to Appendices A and B.

\section{Theory}
Our starting point is the multi-band Hubbard Hamiltonian
\begin{equation}
\label{eq:hamilt}
H=\sum_{\langle \mbR \mbR' \rangle} 
t_{\mbR- \mbR'}^{ij}
\cd_{\mbR i}
\cpd_{\mbR' j} +
\frac{1}{4}\sum_{\mbR}U_{ij,kl}
\cd_{\mbR l}
\cd_{\mbR i}
\cpd_{\mbR j}
\cpd_{\mbR k},
\end{equation} 
where $t_{\mbR- \mbR'}^{ij}$ is the hopping amplitude between orbital $j$ on site
$\mbR'$ and orbital $i$ on site $\mbR$, $\cd_{\mbR i}$ $(c_{\mbR' j})$ are the corresponding
creation (annihilation) operators, and $U_{ij,kl}$ is the anti-symmetrized local Coulomb interaction.
Throughout the text we do not distinguish between spin and
orbital degrees of freedom. Summation over repeated orbital indices is assumed.

The developments and calculations reported here fall within the framework
of dynamical mean-field theory. Detailed discussion of the formalism and basic applications
of DMFT can be found in Ref. \onlinecite{rmp}. The main feature of DMFT is
that the irreducible vertex functions (see below for details) are build only 
from local propagators
and thus can be obtained from an effective quantum impurity problem.
Evaluation of the two-particle response functions can be viewed as a postprocessing 
of the solution to the DMFT equations, which self-consistently determine the fermionic
bath for the effective impurity problem.

\subsection{Linear response formalism}
We review briefly the DMFT linear response formalism. For
details the reader is referred to Ref. \onlinecite{rmp}.
We are interested in the response of a system controlled by (\ref{eq:hamilt}) 
to static fields which couple to the spin, charge or a more general orbital density described
by the susceptibility 
\begin{equation}
\label{eq:susc}
\begin{split}
\chi_{ij,kl}(\mbq)=  
\int_0^{\beta}d\tau 
\sum_{\mbR}e^{i\mbq\mbR} 
\bigl( \la T 
\cd_{\mbR j}(\tau)
\cpd_{\mbR i}(\tau)
\cd_{\mathbf{0} k}(0)
\cpd_{\mathbf{0} l}(0)
 \ra \\
-\la 
\cd_j
\cpd_i
\ra
\la 
\cd_k
\cpd_l
\ra \big).
\end{split}
\end{equation}
Here $\mbq$ is a vector from the first Brillouin zone and the site indices were dropped in the local averages. 
In the DMFT approximation the susceptibility (\ref{eq:susc}) can be obtained from
\begin{equation}
\label{eq:om_sum}
\chi_{ij,kl}(\mbq)=T\sum_{m,n}\tc_{ij,kl}(\mbq;\om_m,\om_n),
\end{equation}
where $\tc_{ij,kl}(\mbq;\om_m,\om_n)$ is the solution of 
coupled integral equations \cite{jarrell92,rmp}
\begin{align}
\begin{split}
\label{eq:bsq}
&\tc_{ij,kl}(\mbq;\om_1,\om_2)=\tc^0_{ij,kl}(\mbq;\om_1,\om_2)\\
&+T\sum_{\om_3,\om_4}\tc^0_{ij,mn}(\mbq;\om_1,\om_3)
\Gamma_{mn,pq}(\om_3,\om_4)\tc_{pq,kl}(\mbq;\om_4,\om_2) 
\end{split}\\
\begin{split}
\label{eq:bsi}
&\tc_{ij,kl}(\om_1,\om_2)=\tc^0_{ij,kl}(\om_1,\om_2)\\
&+ T\sum_{\om_3,\om_4}\tc^0_{ij,mn}(\om_1,\om_3)
\Gamma_{mn,pq}(\om_3,\om_4)\tc_{pq,kl}(\om_4,\om_2).
\end{split}
\end{align}
The local and corresponding 'lattice' quantities are distinguished by the presence
of parameter $\mbq$. The quantities is the equations are functions of 
the discrete fermionic Matsubara frequencies, which are  at temperature $T$
given by $\om_n=(2n-1)\pi T$ ($n$ is an integer).
The equations have the form of the irreducible Bethe-Salpeter (IBS) equation,
with $\Gamma_{mn,pq}(\om_3,\om_4)$ being the particle-hole irreducible vertex at
zero energy transfer,
in the diagrammatic expansion of which the $mn$ pair of external vertices cannot be separated from $pq$ pair by
cutting two single-particle propagators. In the DMFT approximation
only the local diagrams contribute to $\Gamma_{mn,pq}(\om_3,\om_4)$ \cite{zlatic90},
which allows simplifying the IBS equation for the full two-particle correlation function
to equation (\ref{eq:bsq}) in terms of the reduced quantity $\tc_{ij,kl}(\mbq;\om_1,\om_2)$ and k-integrated
particle-hole bubble
\begin{equation}
\label{eq:bubleq}
\tc^0_{ij,kl}(\mbq;\om_1,\om_2)=-\frac{\delta_{\om_1\om_2}}{N}\sum_{\mbk}G_{ik}(\mbk;\om_1)G_{lj}(\mbk+\mbq;\om_1).
\end{equation}
$G_{ik}(\mbk;\om_1)$ is the single-particle propagator obtained from the Dyson equation
\begin{equation}
\label{eq:dyson}
G_{ik}(\mbk;\om_n)=\bigl[i\om_n+\mu-h_{\mbk}-\Sigma(\om_n)]\bigr]^{-1}_{ik},
\end{equation}
where $\mu$ is the chemical potential, $h_{\mbk}$ is the Fourier transform of the single-particle
part of  the Hamiltonian (\ref{eq:hamilt}) and $\Sigma(\om_n)$ is the local self-energy.
Equation (\ref{eq:bsi}) is the IBS equation for the impurity, relating 
the local two-particle correlation function  
\begin{equation}
\label{eq:tpi}
\begin{split}
&\tc_{ij,kl}(\om_1,\om_2)=
T^2\int_0^{\beta} d\tau_1 \int_0^{\beta} d\tau_2 
\int_0^{\beta} d\tau_3 \int_0^{\beta}  d\tau_4 \\
& e^{i\om_1(\tau_1-\tau_2)}e^{i\om_2(\tau_3-\tau_4)}
\bigl( \la T 
\cpd_i(\tau_1)\cd_j(\tau_2)\cpd_l(\tau_3)\cd_k(\tau_4) \ra \\
& -\la T \cpd_i(\tau_1)\cd_j(\tau_2)\ra \la T \cpd_l(\tau_3)\cd_k(\tau_4) \ra \bigr)
\end{split}
\end{equation}
to the local particle-hole bubble
\begin{equation}
\label{eq:bublei}
\tc^0_{ij,kl}(\om_1,\om_2)=-\frac{\delta_{\om_1\om_2}}{N^2}\sum_{\mbk\mbk'}G_{ik}(\mbk;\om_1)G_{lj}(\mbk';\om_1).
\end{equation}

\subsection{Numerical implementation}
Although it is possible to eliminate the vertex function from equations (\ref{eq:bsq}) and (\ref{eq:bsi}),
we evaluate $\Gamma_{ij,kl}(\om_1,\om_2)$ explicitly as an intermediate step. The computation proceeds as follows.
After converging the DMFT equations we use the impurity self-energy $\Sigma_{ij}(\om_n)$ to
get $\tc^0_{ij,kl}(\mbq;\om_1,\om_2)$ and $\tc^0_{ij,kl}(\om_1,\om_2)$
using expressions (\ref{eq:bubleq}), (\ref{eq:bublei}), and (\ref{eq:dyson}).
The local two-particle correlation function $\tc_{ij,kl}(\om_1,\om_2)$ 
is obtained during the solution of the impurity problem. 
Next, equation (\ref{eq:bsi}) is inverted to obtain $\Gamma_{ij,kl}(\om_1,\om_2)$.
Finally, equation (\ref{eq:bsq}) is solved and the susceptibility
$\chi_{ij,kl}(\mbq)$ obtained after summation over the Matsubara frequencies (\ref{eq:om_sum}).

The above program faces two numerical challenges. 
First, the inputs to the IBS equations, $\tc_{ij,kl}(\om_1,\om_2)$ in particular, are known accurately only for a limited
range of frequencies. Second, $\tc_{ij,kl}(\om_1,\om_2)$, $\tc^0_{ij,kl}(\mbq;\om_1,\om_2)$,
$\tc_{ij,kl}(\mbq;\om_1,\om_2)$, and $\tc^0_{ij,kl}(\mbq;\om_1,\om_2)$ decay
as $1/\om^2$, which makes straightforward inversion of (\ref{eq:bsq}) and (\ref{eq:bsi})
numerically unstable. We have been using the Hirsch-Fye quantum Monte-Carlo impurity solver \cite{hirsch86}, but
these issues are general and apply to other methods as well.
Our computational procedure relies on splitting the problem
into a low and a high frequency part, which are treated in numerically different ways.

\subsubsection{Asymptotic behavior of the vertex function}
Separation of low and high frequency parts of a given problem  
is common in theoretical physics. In the DMFT practice  it is widely used
in calculation of the self-energy, where the numerical solution of the
Dyson equation at low-frequencies is augmented by a high-frequency asymptotic
expansion obtained from the moments of the spectral function \cite{oudovenko02}.
In Appendix A we start by considering the HF asymptotic
expansion for the self-energy, which serves a precursor for the HF expansion 
of the vertex function $\Gamma_{ij,kl}(\om_1,\om_2)$.

Unlike the self-energy, where a separate equation exists for each frequency
and so the LF and HF parts are decoupled,
in case of equations (\ref{eq:bsq}) and (\ref{eq:bsi}) we are dealing with matrices in 
the Matsubara frequencies and have to solve coupled equations for the HF and LF blocks. 
The key ingredient of our procedure is replacing the vertex function
outside the LF block by its asymptotic form
\begin{equation}
\label{eq:gainf}
\begin{split}
&\gi_{ij,kl}(\om_1,\om_2)= U_{ij,kl}+U_{im,kn}\chi^{ph}_{mn,pq}(\om_1-\om_2)U_{pj,ql} \\
                           &+\frac{1}{4}U_{im,nl}\chi^{pp}_{mn,pq}(\om_1+\om_2)U_{pj,kq} \\ 
&\quad\text{for}\quad |\om_1|>\om_c\vee|\om_2|>\om_c,
\end{split}
\end{equation}
where $\chi^{ph}(\nu)$ and $\chi^{pp}(\nu)$ are the local dynamical susceptibilities 
in the particle-hole and particle-particle channel, respectively,  as functions of bosonic Matsubara 
frequency $\nu$, 
and $\om_c$ is the cut-off frequency defining the LF block.
\begin{align}
\begin{split}
\chi^{ph}_{ij,kl}(\nu)&=\int_0^{\beta}d\tau \exp(i\nu\tau)\bigl(\la T 
\cd_j(\tau)
\cpd_i(\tau)
\cd_k(0)
\cpd_l(0) \ra \\ &
-\la 
\cd_j 
\cpd_i\ra 
\la 
\cd_k 
\cpd_l\ra\bigr) 
\end{split}
\\
\chi^{pp}_{ij,kl}(\nu)&=\int_0^{\beta}d\tau \exp(i\nu\tau)
\la T 
\cpd_i(\tau)
\cpd_j(\tau)
\cd_k(0)
\cd_l(0)\ra
\end{align}

Derivation of expression (\ref{eq:gainf})is given in Appendix A.
Viewed as a function of variables $\om_1$ and $\om_2$, $\gi_{ij,kl}$ 
is a constant, $U_{ij,kl}$, plus two ridges along the main and the minor diagonal, a structure 
observed in earlier numerical studies \cite{toschi07}. The cross section of these ridges is 
given by the local dynamical susceptibility, typically, with a fast decay away
from $\nu=0$. The asymptotic form (\ref{eq:gainf}) is used in the numerical treatment
of both equations (\ref{eq:bsq},\ref{eq:bsi}) as described below.

\subsubsection{Blocked IBS equations}
The relations (\ref{eq:bsq},\ref{eq:bsi}) can be viewed as equations between matrices indexed
by the Matsubara frequencies and pairs of the orbital indices. 
We use the following arrangement of the matrices  
\begin{equation*}
\begin{pmatrix}
&\mbc(-\om_1,-\om_1)&\mbc(-\om_1,+\om_1)&\quad\mbc(-\om_1,-\om_2)&\cdots \\
&\mbc(+\om_1,-\om_1)&\mbc(+\om_1,+\om_1)&\quad\mbc(+\om_1,-\om_2)&\cdots \\
& & & & \\
&\mbc(-\om_2,-\om_1)&\mbc(-\om_2,+\om_1)&\quad\mbc(-\om_2,-\om_2)&\cdots \\
&\vdots &\vdots &\vdots &
\end{pmatrix},
\end{equation*}
where each element $\mbc$ is itself a matrix in pairs of orbital indices
\begin{equation*}
\mbc=\!
\bordermatrix{
  & 11 & 21 & \cdots & 12 & \cdots  \cr
11& \chi_{11,11} & \chi_{11,21} &\cdots & \chi_{11,12} & \cdots \cr
21& \chi_{21,11} & \chi_{21,21} &\cdots & \chi_{21,12} & \cdots \cr
~\vdots &\vdots&\vdots&\ddots&\vdots& \cr
12 & \chi_{12,11} & \chi_{12,21} &\cdots & \chi_{12,12} & \cdots \cr
~\vdots &\vdots&\vdots&&\vdots& \ddots \cr
}.
\end{equation*}
In this representation the LF block is located in the upper left corner of the 
matrix and the $\om$ dependent part of $\gi_{ij,kl}(\om_1,\om_2)$ (\ref{eq:gainf}) 
proportional to $\chi^{ph}$ and $\chi^{pp}$ has a band diagonal form.
 
In the following we split the set of Matsubara frequencies into the LF block $|\om_n|<\om_c$,
denoted with a block index 0, and HF block $|\om_n|\ge\om_c$, denoted with a block index 1.
Expansion of a matrix equation of the form (\ref{eq:bsq},\ref{eq:bsi}) into the LF and HF blocks
is given in Appendix B. 
The impurity IBS equation (\ref{eq:bsi}) is used to compute the LF part of 
the vertex function, the $\Gamma^{00}$ block.
We assume to know the LF (00) part of $\tc_{ij,kl}(\om_1,\om_2)$ (\ref{eq:tpi}) from the solution of the
impurity problem
and to have a complete information about $\tc^{0}_{ij,kl}(\om_1,\om_2)$ (\ref{eq:bublei}). 
In practice we use a numerical representation of $\tc^{0}_{ij,kl}(\om_1,\om_2)$ up to a 
HF cut-off $\Omega_c$ (typically much larger that the LF cut-off $\om_c$)
and the analytic $-\frac{1}{\omega_n^2}$ tail up to infinity for elements the $\tc^{0}_{ij,ij}$ elements, 
while the remaining elements involving off-diagonal $G_{ij}$ (\ref{eq:bublei})
are neglected above $\Omega_c$. 
The LF part of the vertex $\Gamma^{00}$ is obtained from (\ref{eq:g00}), 
which amounts to inverting equation (\ref{eq:bsi})
restricted to the LF block and subtracting a correction.
For $\chi^{ph}(\nu)$ and $\chi^{pp}(\nu)$ sharply peaked around $\nu=0$ the correction
reduces to a constant shift everywhere except the vicinity of $\om_c$.
\begin{figure}
\includegraphics[angle=270,width=0.9\columnwidth,clip]{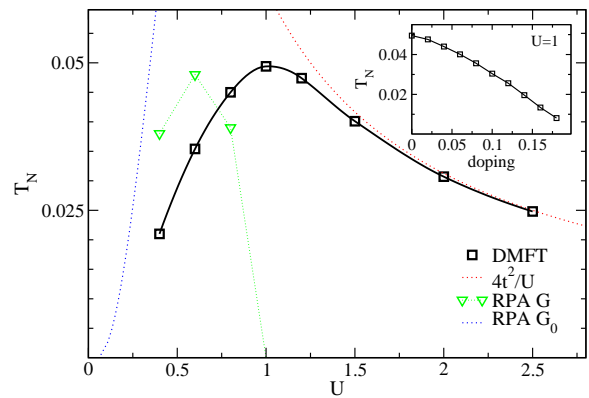}
\caption{\label{fig:tn}The N\'eel temperature of the 2D Hubbard model in DMFT approximation (squares),
shown together with the mean-field value of the Heisenberg model (red, large U) and
the RPA results obtained with the non-interacting (dotted blue line) and dressed (triangles)
propagators. The inset shows the decrease of the N\'eel temperature upon doping away
from half-filling at $U=1$.}
\end{figure}

Once we have computed $\Gamma^{00}$ we proceed to solve equation (\ref{eq:bsq})
for $\tc_{ij,kl}(\mbq;\om_1,\om_2)$ assuming the full knowledge of
$\tc^0_{ij,kl}(\mbq;\om_1,\om_2)$ and $\Gamma_{ij,kl}(\om_1,\om_2)$.
We use the same partitioning into the LF and HF blocks and 
take advantage of the fact that only $\om$-summed
susceptibility $\chi_{ij,kl}(\mbq)$ is of interest. This allows partial $\om$-summations
in the HF block to be preformed at an earlier stage of the calculation and to replace
matrix-matrix operations with matrix-vector ones, see Appendix B. 
Further speed-up comes from treating the $\om$-dependent (\ref{eq:band}) and the constant (\ref{eq:const})
parts of  $\Gamma^{11}$ in two subsequent steps, which reduces the
most demanding part of the HF problem to a repeated band-matrix--vector multiplication.

\begin{figure}
\centering\includegraphics[width=0.49\columnwidth,clip]{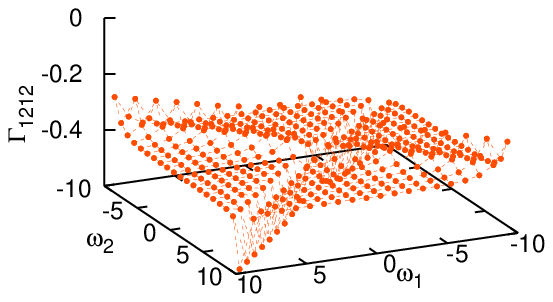}
\includegraphics[width=0.49\columnwidth,clip]{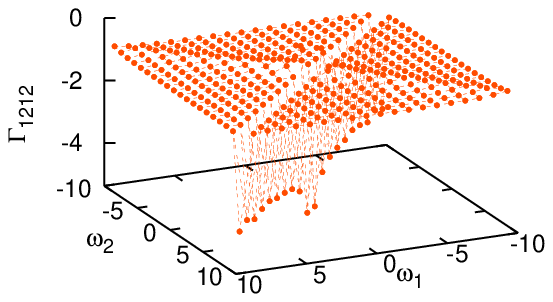}
\caption{\label{fig:ga} The real part of the transverse spin vertex $\Gamma_{1212}(\om_m,\om_n)$ as a 
function of the indices $m$, $n$ for $U=0.4$ (left) and 1 (right)
obtained at $T=1/15$.}
\end{figure}
\begin{figure}
\includegraphics[angle=270,width=0.9\columnwidth,clip]{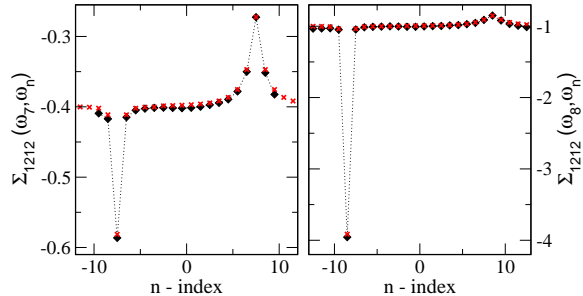}
\caption{\label{fig:vert_match} The real part of the transverse spin vertex $\Gamma_{1212}(\om_m,\om_n)$ 
for fixed $\om_m$ in the asymptotic range for $U=0.4$ (left) and 1 (right)
at $T=1/15$. Full diamonds mark the numerical data from LF block, the crosses denote the
HF tail obtained from expression \ref{eq:gainf}.}
\end{figure}

\section{Numerical results}
\subsection{2D Hubbard model}
In the following we present our results for 2D Hubbard model on a square lattice with the nearest
neighbor hopping. In particular, we compute the N\'eel temperature as a function of U and doping.
This is an academic example since it is well know that no magnetic order is possible in this model 
at finite temperature and the magnetic oder studied here is an artefact of the DMFT approximation.
Nevertheless, we find it a useful benchmark since i) a comparison to other studied is possible and
ii) the model exhibits a crossover from a nesting-driven instability at small $U$ to 
the Heisenberg model with local moments at large $U$. 

First, we report the results obtained at half filling. 
After converging the DMFT equations, a finer QMC run of $5\times10^6$ to
$10^7$ sweeps is performed to obtain the local two-particle correlation function
$\tc_{ij,kl}(\om_1,\om_2)$ and local susceptibilities $\chi^{ph}_{ij,kl}(\nu)$, $\chi^{pp}_{ij,kl}(\nu)$. 
It is possible
to separate the charge and the spin-longitudinal channels from the beginning, 
but we perform the calculation in the upup, updn, dnup, dndn basis, as would be the case for general 
orbital indices, and separate the two channels only in the final result. 
Expression (\ref{eq:bubleq}) is evaluated on a $20\times20$
grid of q-points (66 irreducible points) using a fine k-mesh 
of $551\times 551$ points, which is more than enough to ensure that the uncertainties associated 
with k-points sampling as much smaller that other sources of error. 
\begin{figure}
\centering\includegraphics[width=0.49\columnwidth,clip]{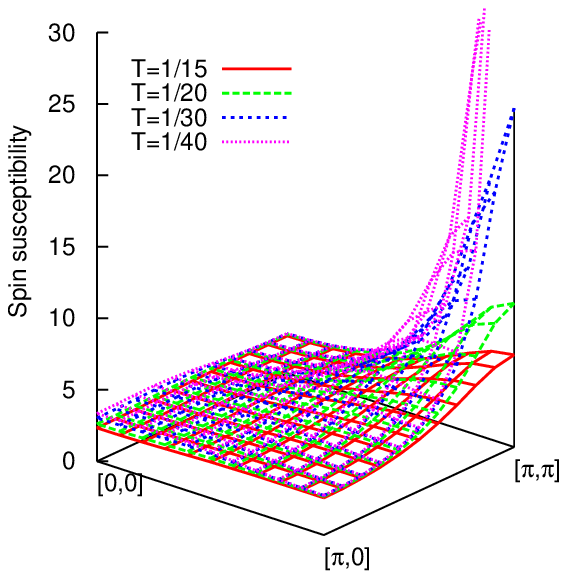}
\includegraphics[width=0.49\columnwidth,clip]{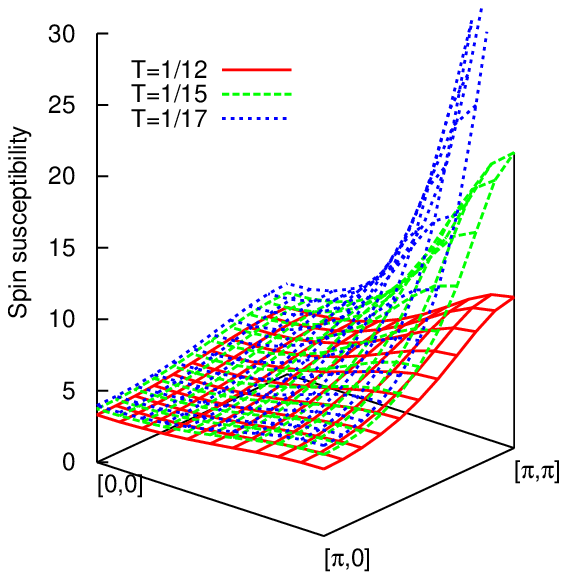}
\caption{\label{fig:ch_spin} The longitudinal spin susceptibility as a function of $\mbq$
at various temperatures for $U=0.4$ (left) and $U=1$ (right).}
\end{figure}  

The IBS equations were solved for several LF cut-offs $\om_c$ leaving between 5 to 20 lowest Matsubara frequencies
for temperatures above 1/20, and 15 to 30 Matsubara frequencies for lower temperatures. 
In all studies cases, the relative variation of the susceptibility with the LF cut-off
was less than 1\%. We have also tested the sensitivity to the HF cut-off $\Omega_c$. 
When neglecting the contributions from above $\Omega_c$ completely, a slow
convergence of the susceptibilities close to the AFM instability
was observed. The error essentially followed the $1/\Omega_c$ dependence 
which may be expected when truncating a sum over $1/\om_n^2$ series. The dominant contribution
to the error came from the $\Gamma^{01}X^{11}\Gamma^{10}$ term in equation (\ref{eq:a00}).
Introduction of an analytic summation of the $1/\om_n^2$ tails to infinity
lead to the converged results already for lowest HF cut-offs corresponding to 80-100 lowest Matsubara frequencies.

To check the consistency of our calculations we have solved equation (\ref{eq:bsq})
also for the local bubble $\tc^0_{ij,kl}(\om_1,\om_2)$, i.e. went from 
$\tc_{ij,kl}(\om_1,\om_2)$ to $\Gamma_{ij,kl}(\om_1,\om_2)$ and back, and compared
the result with the local susceptibility obtained directly from the QMC simulation.
For all reported parameters we have found a relative deviations of less than
2\% close to the AFM instability and less than 0.5\% deeper in the paramagnetic phase.
As a side remark we mention that in the insulating phase for $U>1$ it is 
important that $\tc^0_{ij,kl}(\om_1,\om_2)$ entering equations (\ref{eq:bsq}) and ((\ref{eq:bsi})
are exactly the same (e.g. obtained with the same k-point sampling) since in this parameter range
the results are extremely sensitive to differences between $\tc^0$'s entering
equation (\ref{eq:bsq}) and (\ref{eq:bsi}).

Due to the SU(2) symmetry of the model the transverse and the longitudinal 
spin susceptibilities are identical in principle. This symmetry is not explicitly enforced in the
QMC calculation, where the identity is obeyed only approximately within the stochastic error 
of the simulation.
For $U<1$ we have found relative deviations between the transverse and longitudinal
spin susceptibilities of 0.2-0.5\%. For larger $U$'s (especially at lower temperatures)
the transverse susceptibility rapidly becomes useless due to increasingly large
errorbars. This problem is specific to the Hirsch-Fye QMC solver and the way 
the transverse susceptibility is measured in the simulation.
We have used the longitudinal susceptibility to determine $T_N$ in this parameter range.

In Fig. \ref{fig:tn} we show the N\'eel temperature as a function of U. On the large U side $T_N$
approaches the static mean field value of the corresponding Heisenberg model reflecting the
insulator with well defined local moments. On the small U side the magnetic order arises from the
Fermi surface nesting. For comparison, we show also $T_N$ for the RPA condition for the
instability, $\chi^0(T_N)=1/U$, obtained by replacing the vertex with a constant 
and taking $\chi^0(T)$ of the non-interacting electron gas or $\chi^0(T)$ obtained with the dressed propagators.

In Fig. \ref{fig:ga} we show the LF block of the real part of the transverse spin vertex $\Gamma_{12,12}$ for large and small $U$ as a
function of the fermionic Matsubara frequencies, which exhibits the two-ridge structure and
a rapid convergence towards its asymptotic form. At small $U$ the ridges are only weakly $T$-dependent and
their absolute value is small compared to the constant part of the vertex. 
At large $U$s the ridges grow as $1/T$ and thus become dominant at low enough temperatures.  
In Fig. \ref{fig:vert_match} we show the same vertices at fixed $\om_m$ (still in the LF block)
together with the corresponding asymptotic tails $\gi_{12,12}$. We find a good match between the two 
dependencies. The remaining mismatch is mainly due to the time discretization error inherent
to the Hirsch-Fye impurity solver, which shows up in a spurious curvature of the $\Gamma(\om_m,\om_n)$
for larger $\om$s (not really visible in the present plots). Using the semi-analytic asymptotic tail $\gi$ 
can efficiently remove this spurious curvature.
The discretization error arises from performing Greens function convolutions on a discrete
time grid and should not be confused with the so called Trotter error, which is another
consequence of discretizing the imaginary time and which cannot be removed be restricting
oneself to the low frequencies. Impurity solvers using continuous time approaches are not
expected to have this problem.

Finally, the spin susceptibility throughout the Brillouin zone
is shown in Fig. \ref{fig:ch_spin}. The plots clearly show that the magnetic instability
takes place at $(\pi,\pi)$ as expected. The behavior of the vertices reveals that the
origin of the instability is quite different for small and large $U$s. 
While for large $U$s the main $T$-dependence in the problem is in the vertex,
related to $1/T$ behavior of the local susceptibility, for small $U$x the main
$T$-dependence is in the 'bubble', namely the of thermal smearing of the nesting related peak.

\subsection{Effect of doping}
\begin{figure}
\includegraphics[angle=270,width=0.9\columnwidth,clip]{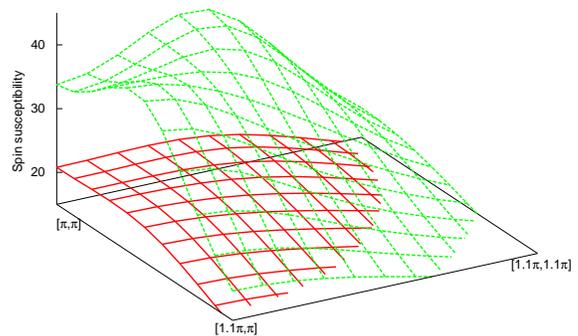}
\caption{\label{fig:018}The detail of the q-dependent spin susceptibility in the vicinity of $[\pi,\pi]$
at $T=1/60$ (larger values) and $T=1/30$ for the doping of 0.18.}
\end{figure}

We use doping to break the particle-hole symmetry in the above model. The calculations were performed 
in the crossover regime between the metallic and insulating phases at $U=1$. 
We have found that magnetic instability
survives doping up to about 0.2 electron (or hole) per site. For dopings less or equal to 
0.14 the instability appears
at $[\pi,\pi]$. For dopings of 0.16 and 0.18 we have noticed that
the maximum of $\chi_s(\mbq)$ moved slightly away from $[\pi,\pi]$ as the temperature
was lowered. Similar behavior was observed also in a model with the next-nearest-neighbor
(diagonal) hopping \cite{toschi_private}.
The N\'eel temperature as a function of doping is shown 
in the inset of Fig. \ref{fig:tn}. In Fig. \ref{fig:018} we show the detail
of the spin susceptibility in the vicinity of $[\pi,\pi]$ for the highest studied
doping of 0.18. While at high temperatures
the susceptibility exhibits a flat maximum at $[\pi,\pi]$, at lower temperatures
the maximum moves to $[1.04\pi,\pi]$ eventually giving rise to an incommensurate order.

\subsection{Bilayer model}
As a last example we study a bilayer Hubbard model \cite{sentef09}. 
We choose this model because: i) it has two bands and the lattice bubble
$\tc_{ij,kl}(\mbq;\om_1,\om_2)$ has a more complicated orbital
structure than for the single-band model
ii) it exhibits a non-trivial temperature dependence of the 
uniform spin susceptibility characterized by an exponential decrease
at low temperatures.
In our previous work on a more general version of this model (model of covalent
insulator) \cite{kunes08} we have used a different numerical approach to
compute the uniform susceptibility without an explicit use of the
vertex function and its HF tail, an approach which could not be used for
more general models and $q$-dependent susceptibilities. Obtaining the exponential decay at low
temperatures proved then numerically challenging. Therefore 
we find this model to provide a good test of 
the performance of the present method. 

The model is obtained by taking two Hubbard sheets from the previous section
(with opposite signs of the in-plane hopping) and adding 
an inter-plane hopping of 0.2. The different signs of the in-plane hoppings 
on the two sheets ensures that any non-zero inter-plane hopping opens a 
hybridization band gap. The inter-plane hopping was chosen such that the interesting
temperature range can be easily covered with the QMC simulations.
As in the other cases studied so far we were not primarily interested in 
the physics of the model, but in the performance of the computational method.
Therefore we omit that fact that the system actually 
exhibits an AFM instability and focus on the uniform susceptibility only.
While the AFM instability can be removed by introducing an in-plane frustration,
the $T$-dependence of the uniform susceptibility is a generic feature
independent of the bandstructure details \cite{kunes08}. 

\begin{figure}
\includegraphics[angle=270,width=\columnwidth]{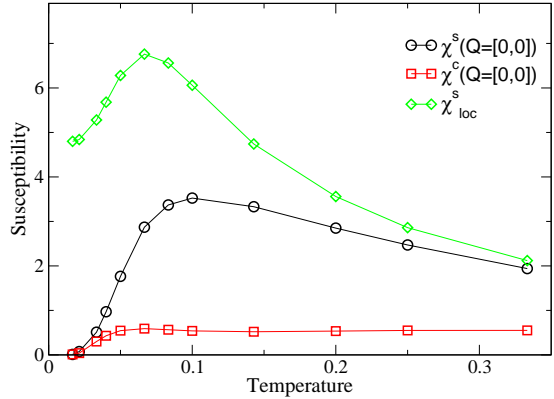}
\caption{\label{fig:bilayer} The uniform spin (circles) and charge (squares) susceptibility
of the bilayer model as a function of temperature. The diamonds mark the local spin susceptibility,
which clearly shows the evolution of local moment above $T=0.1$.}
\end{figure}

In Fig. \ref{fig:bilayer} we show the local and uniform spin susceptibilities
together with the uniform charge susceptibility, which is inversely proportional
to the compressibility of the electron liquid. The calculated spin susceptibilities
closely resemble those of Ref. \onlinecite{kunes08}. The decay of the uniform susceptibilities
at low temperatures reflects an evolution towards a band-insulator-like state,
characterized by a finite excitation gap for both charge and spin excitations
(for more information see Ref. \onlinecite{sentef09}).
In the present calculations we have not encountered any numerical problems in
reproducing the low values of the susceptibility at low $T$.

\section{Conclusions}
An efficient numerical procedure for calculation of the static two-particle response
functions within the DMFT formalism was presented. The main development is
the identification of the asymptotic form of the two-particle irreducible vertex function
and implementation of a block algorithm for solution of the irreducible
Bethe-Salpeter equations. 
We have studied several test cases including the
single-band Hubbard model at and away from half-filling, and a bilayer
Hubbard model at half-filling using Hirsch-Fye quantum Monte-Carlo impurity solver.
In all cases we have observed a rapid convergence of the vertex
function towards its asymptotic form and a very good numerical stability of the computations. 
This means that the numerical effort can
be efficiently reduced by choosing a relatively small low-frequency cut-off,
which makes calculations for multi-band systems feasible.
Besides the susceptibility calculations the described procedure
may be useful in diagrammatic extensions of DMFT, which employ the IBS formalism
and two-particle vertices \cite{kusunose06,rubtsov08,toschi07,slezak09}.

\begin{acknowledgments}
I'd like to thank M. Jarrell for sharing his ideas about  
the high frequency part of the Bethe-Salpeter equation, A. Toschi for discussing his
unpublished results on the incommensurate order in the doped Hubbard model,
A. \v{C}ejchan for help with symbolic calculations at the early stage of this work,
K. Byczuk, D. Vollhardt, A. Kampf,
A. Kauch, K. Held and V. Jani\v{s} for discussions at various stages of this work.
This work was supported by grants no. P204/10/0284 and 202/08/0541 of the Grant Agency 
of the Czech Republic and by the Deutsche Forschungsgemeinschaft through FOR 1346.
\end{acknowledgments}

\appendix
\section{High frequency expansion of the vertex function}
Following Baym and Kadanoff \cite{BK} and Baym \cite{baym62} the self-energy $\Sigma$
can be viewed as a functional of the dressed Green function $G$ and the vertex function
$\Gamma$ can be obtained as a variation $\tfrac{\delta \Sigma[G]}{\delta G}$. 
The self-energy itself can be obtained as a variation of the Baym-Kadanoff functional
$\Sigma=\tfrac{\delta \Phi[G]}{\delta G}$. Diagrammatically this means that 
$\Sigma$ is obtained from $\Phi$ by cutting a single Green function line and
$\Gamma=\tfrac{\delta^2 \Phi[G]}{\delta G \delta G}$ is obtained by cutting two 
Green function lines. 
\begin{figure}
\includegraphics[width=0.66\columnwidth,clip]{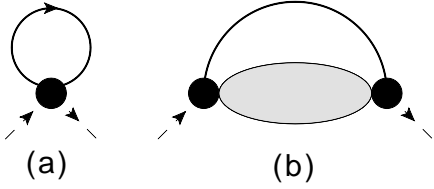}
\caption{\label{fig:sig2} The form of the diagrams for the self-energy $\Sigma(\om)$, which 
contribute to the zero's (a) and first (b) order in $\tfrac{1}{\om}$.}
\end{figure}

We are interested in the high frequency behavior of $\Sigma(\om)$ and $\Gamma(\om_1,\om_2)$.
The expansion of $\Sigma(\om)$ to the order $\tfrac{1}{\om}$ is well known
from the moments of the spectral function \cite{oudovenko02}. 
Diagrammatically the zero's order term of $\Sigma$ is the
Hartree-Fock contribution, in which the external frequency does not enter any of
the internal lines (Fig. \ref{fig:sig2}a). 
The first order contribution, $1/{\om}$, comes from diagrams with 
external vertices connected by single Green function line (Fig. \ref{fig:sig2}b).
Their contribution has the form
\begin{equation}
\label{eq:a1}
T\sum_{\nu}U\chi(\nu)UG(\pm\om-\nu)
\end{equation}
with $\chi(\nu)$ being the dynamical susceptibility,
where the plus sign in front of $\om$ applies for particle-hole and the negative sign
for particle-particle susceptibility.
Expanding $G$ in (\ref{eq:a1}) in the powers of $\tfrac{1}{(\om-\nu)}=\tfrac{1}{\om}+\tfrac{\nu}{\om(\om-\nu)}$ and
taking the leading $\tfrac{1}{\om}$ term we get the corresponding contribution 
to the self-energy. The remaining free summation over $\nu$ leads to 
an equal-time correlator of the $\la \cd \cpd \cd \cpd \ra$ type.

\begin{figure}
\includegraphics[width=\columnwidth]{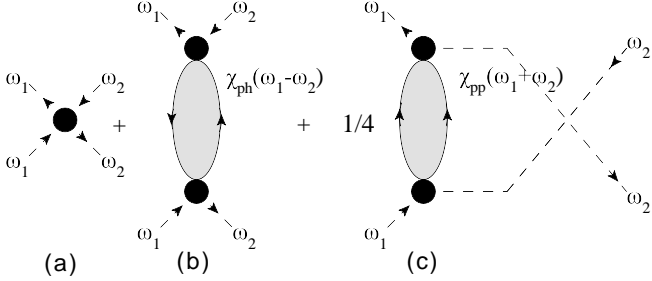}
\caption{\label{fig:vertex} The diagrams for the irreducible vertex, which remain non-zero in the
limit of large $\om_1$ and $\om_2$.}
\end{figure}

Our main interest here lies in the high frequency behavior of $\Gamma(\om_1,\om_2)$. In particular,
we are interested in the zero's order contribution, which remains finite in the
limit or large $\om_1$ and/or $\om_2$. It is not possible to develop the high frequency
expansion by keeping one of the frequencies fixed since the limit $\om_1,\om_2\rightarrow\infty$
is not uniform, i.e. it depends on the way in which infinity is approached.
By inspecting the diagrams for $\Gamma(\om_1,\om_2)$ we have identified two
non-vanishing contributions. The first one, obtained from the variation of the 
zero's order term in the self-energy (Fig. \ref{fig:sig2}a) is simply the 
bare vertex Fig. \ref{fig:vertex}a. The other non-vanishing terms are obtained
by cutting the Green function line in Fig. \ref{fig:sig2}b, which leads
to Fig. \ref{fig:vertex}b and Fig. \ref{fig:vertex}c. The corresponding
expression for $\gi$ reads
The expression $\gi$ reads
\begin{equation}
\begin{split}
\gi_{ij,kl}(\om_1,\om_2)=& U_{ij,kl}+U_{im,kn}\chi^{ph}_{mn,pq}(\om_1-\om_2)U_{pj,ql} \\
                           & +\frac{1}{4}U_{im,nl}\chi^{pp}_{mn,pq}(\om_1+\om_2)U_{pj,kq}.
\end{split}
\end{equation}
The factor $1/4$ accounts for the exchanges of the external points of the two vertices.
Note, that the second order diagram (obtained by replacing the susceptibility in Fig. \ref{fig:sig2}b
with a simple bubble) contributes to both Fig. \ref{fig:vertex}b and Fig. \ref{fig:vertex}c.
As a function of $\om_1$ and $\om_2$ the high frequency part of the vertex function 
$\gi$ has a structure of a constant plus two ridges along the main and minor diagonals,
cross section of which is determined by the particle-hole and particle-particle dynamical susceptibilities.
\begin{figure}
\includegraphics[width=0.85\columnwidth]{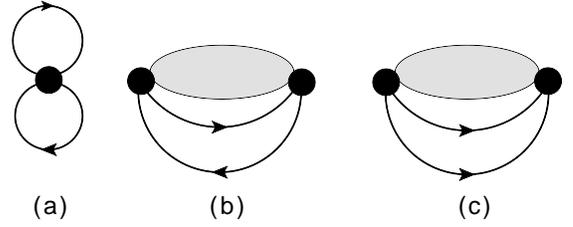}
\caption{\label{fig:bk} The diagrams in the Baym-Kadanoff functional $\Phi[G]$, which contribute
to $\gi$.}
\end{figure}

In terms of the Baym-Kadanoff functional, the constant term in $\gi$ derives from the first order, Hartree-Fock, diagram
(Fig. \ref{fig:bk}a),
and the diagrams containing at least one particle-hole or particle-particle
bubble (Fig. \ref{fig:bk}b,c) by cutting the Green function lines corresponding to the same bubble. 

\section{Blocked IBS equations}
In the following we present the IBS equation written in terms of small-$\om$ and large-$\om$ 
sectors denoted by indices 0 and 1, respectively. The IBS equation has the form
\begin{equation}
A=B+B\Gamma A,
\end{equation}
where $A$, $B$ and $\Gamma$ are matrices indexed by the discrete Matsubara frequencies. Matrix
B is diagonal in the Matsubara frequencies.

In the first step we assume that $A^{00}$, $B^{00}$, $B^{11}$, $\Gamma^{11}$, $\Gamma^{01}$,
and $\Gamma^{10}$ are known and we want to compute $\Gamma^{00}$. 
From
\begin{align}
A^{00}=B^{00}+&B^{00}\Gamma^{00}A^{00}+B^{00}\Gamma^{01}A^{10} \\
A^{10}=&B^{11}\Gamma^{10}A^{00}+B^{11}\Gamma^{11}A^{10}
\end{align}
we obtain
\begin{equation}
\label{eq:g00}
A^{00}=B^{00}+B^{00}\bigl(\Gamma^{00}+\Gamma^{01}X^{11}\Gamma^{10}\bigr)A^{00},
\end{equation}
where $X^{11}$ full-fils the equation
\begin{equation}
\label{eq:x11}
X^{11}=B^{11}+B^{11}\Gamma^{11}X^{11}.
\end{equation}
Thus $\Gamma^{00}$ can be obtained by inverting IBS equation truncated to the 00 block and adding
a correction term.

In the second step we assume full knowledge of $\Gamma$ and $B$ and want to compute $A$. Importantly,
we are interested only in the sum over the matrix elements (Matsubara frequencies) of $A$.
To this end we use a set of block equations
\begin{align}
\label{eq:a00}
A^{00}&=B^{00}+B^{00}\bigl(\Gamma^{00}+\Gamma^{01}X^{11}\Gamma^{10}\bigr)A^{00}\\
\label{eq:a10}
A^{10}&=X^{11}\Gamma^{10}A^{00}\\
\label{eq:a01}
A^{01}&=A^{00}\Gamma^{01}X^{11}\\
\label{eq:a11}
A^{11}&=X^{11}+X^{11}\Gamma^{10}A^{00}\Gamma^{01}X^{11}.
\end{align}
The advantage of the blocked scheme is that equation (\ref{eq:x11}) does not have to be inverted
in practice. This is thanks to three factors:
i) only sum of elements of $X^{11}$ (or a weighted sum in case of $X^{11}\Gamma^{10}$) is
needed, ii) $B^{11}$ is small due to $1/\om_n^2$ decay, iii) $\Gamma^{11}$ is a constant plus
a band matrix. 
We describe in detail the computation of
\begin{equation}
X_{\bullet\bullet}=\sum_{i,j}X_{ij}.
\end{equation}
The computation for $X^{11}\Gamma^{10}$ is analogous.
First, we solve iteratively equation (\ref{eq:x11}) taking only the band part of $\Gamma^{11}$
\begin{equation}
\label{eq:band}
\tilde{X}_{i\bullet}=B_{ii}+B_{ii}\sum_k \Gamma^{\text{band}}_{ik}\tilde{X}_{k\bullet},
\end{equation}
where the block indices 11 were dropped for simplicity. After $\tilde{X}_{i\bullet}$
is computed the summation over the remaining index is performed to obtain $\tilde{X}_{\bullet\bullet}$.
Next, we should solve equation (\ref{eq:x11}) taking the constant part of $\Gamma^{11}$ and
$\tilde{X}$ in place of $B^{11}$. This is not necessary, since there is a simple, well known,
relationship between ${X}_{\bullet\bullet}$ and $\tilde{X}_{\bullet\bullet}$
\begin{equation}
\label{eq:const}
X_{\bullet\bullet}=\bigl(I-\tilde{X}_{\bullet\bullet}\Gamma^{\text{const}}\bigr)^{-1}\tilde{X}_{\bullet\bullet}.
\end{equation}
Note, that quantities in this equation should be viewed as non-commuting matrices in the orbital indices.

\bibliography{v2}
\end{document}